\documentstyle[12pt]{article}

\title{Heterogeneous condensation in dense media}

\author{V.Kurasov}

\begin{document}

\maketitle

\begin{abstract}
The theoretical description of the heterogeneous nucleation kinetics is
presented. This description takes into account the perturbation of the
vapor phase initiated by the growing droplets. The form of the density
profile around the growing droplet is analyzed which leads to some special
approximations. Then the process of  nucleation in the whole system is
described. As the result all main characteristics of the process are
determined
analytically.

\end{abstract}

Among numerous examples of the first order phase transitions
a case of condensation is extracted by  a relative simplicity.
This case is well investigated experimentally and
is traditionally regarded as the base for application of some
new theoretical methods.
A "classical theory of condensation" (see, for example \cite{class}) gives
the solid ground for further theoretical constructions. Numerous modifications
and reconsiderations (see, for example \cite{nonclass}) allow to consider
an example of condensation as well analyzed both theoretically and
experimentally.

One has to stress that practically all investigations were intended to
determine the  rate of nucleation and didn't give the global picture of
the phase transition.
Some theoretical descriptions of a global evolution appeared later than
the classical theory of nucleation has been created and they were not so numerous.
Among them one can extract  the descriptions of  a metastable phase decay
by Wakeshima \cite{Waka}, Segal' \cite{Segal} and Kuni, Grinin
\cite{Kuni-Grinin}.
A process of condensation which occurs during a smooth variation of external
conditions was considered in  \cite{PhysRevE}.
Nevertheless the theoretical description of the global picture of a whole
condensation
kinetics have ignored some important features of this process.
Namely, an exhaustion of a metastable phase near a growing embryo of a new
phase
have not been taken into account in a proper way.
Certainly, this exhaustion is explicitly taken into account in expression
for the rate of embryos growth in continuous  model  (i.e.  in  the
"diffusion
regime" of the embryos growth).  This effect was carefully analyzed both
in stationary and non-stationary aspects in many publications mainly in
the field of mechanics of continuous media.
But the presence of the gap of a metastable phase density
near a droplet
will act on the intensity of the new droplets formation.
 This effect hasn't been taken into account in all previous theoretical
descriptions of the global evolution in the first order phase transition.
But as it is shown in \cite{PhysA} this gap can lead to big numerical
effects in the description of the whole process.

The reason why the mentioned effect was not considered is rather trivial.
Even under a  spatial homogeneous   consumption of the metastable phase the
mentioned
descriptions were rather complicate \cite{PhysRevE}. Then
ordinary the condensation
process was described under the free molecular regime of growth where there
is will be no such a gap.
This was a serious restriction of the theoretical description.

In some publications
(see, for example \cite{Smir}) the regime of the droplets growth was
the diffusion one. This requires to consider the gap of the density
near the growing droplet but in \cite{Smir} the vapor consumption was regarded
to be  homogeneous in space.   As far as this effect is very
important one can not present a reliable description without taking it
into account.
Here we shall present a more realistic picture of a phase transition
which allows an analytical solution.

Qualitatively the picture of the condensation process is rather simple.
A process of nucleation (i.e. a formation of supercritical embryos of
liquid phase) leads to a vapor exhaustion which stops the nucleation process.
But the supercritical embryos continue to consume a vapor phase. All surplus
substance of a metastable phase will be accumulated in the embryos of a
new phase. One can say that the process of condensation is
completed\footnote{A
further evolution  includes the consumption of some relatively small embryos
by some relatively big ones. It will be seen later that
when all surplus substance is consumed then
all droplets
have approximately one and same size
and we don't analyze this process here.}.

The global picture of homogeneous condensation with explicit account of
the density profiles was presented in \cite{PhysA} where the giant
numerical effects were observed. But ordinary the process of nucleation
occurs on  heterogeneous centers\footnote{Also it is more simple to observe
heterogeneous case experimentally.}. This radically complicates the
theoretical
description due to the centers exhaustion which has essentially non-linear
character.

During the nucleation process  one can see that heterogeneous centers
become the centers of the supercritical droplets which are growing
irreversibly
in time. But the nucleation process diminishes the number of free
(unoccupied by droplets)
heterogeneous
centers. In some cases the total exhaustion of free heterogeneous centers
interrupts the nucleation, in some cases the partial exhaustion seriously
influences upon the nucleation rate. This effect has to be also taken
into account.

A simple analytical description of heterogeneous condensation
will be presented here with proper account of all mentioned problems.
As the result all main characteristics of the condensation process will
be expressed through the  parameters of external conditions behavior
and substance parameters by some explicit analytical formulas. The error of
the presented description will be estimated.

The structure of the theory will be as follows:
\begin{itemize}

\item
At first we shall analyze the density profile around the solitary droplet
and construct some approximations. It has much in common with the case
of homogeneous condensation considered in \cite{PhysA} and will be considered
briefly.

\item
Then we shall construct some models for kinetics of process. We have to
show that these models estimate the time evolution of the system during
the nucleation period. As far as these models will give the similar results
one can state that the approximate description of the nucleation kinetics
is given. The error of description is, thus, estimated.

\item
When the solution is already obtained we can compare it with formulas given
by the previous approach  and see the numerical effect
of account of the gap near the growing droplets.

\end{itemize}

The small parameter of the theory will be the
inverse number of  molecules inside
the critical embryo of a new phase. The small  value of this parameter isn't
a restriction of our theory - it  comes  from  the  possibility  of
thermodynamic
approach to calculate the free energy of the critical embryo.
There is no other reliable way to  calculate  the  free  energy  except  the
thermodynamic
approach\footnote{All microscopic models requires very complex calculations
which can not be fulfilled directly.}. To use the thermodynamic approach it
is necessary to have at least a few dozens of molecules inside the embryo.

Also we shall require the barrier character of nucleation. It means that
to begin to grow irreversibly every embryo has to overcome the activation
barrier of essential height. This height is less than for the homogeneously
(pure fluctuation) formed embryo but still attains several
thermal units\footnote{All
energy-like values will be measured in thermal units.}. Certainly, one
can imagine the situation when there is no activation barrier. Then all
embryos begin immediately  to  grow  irreversibly,  The  number  of
droplets
(i.e. the irreversibly growing embryos) will be equal to the total number
of centers and the kinetics of the process will be relatively simple.

We shall speak only about the density profiles around the droplet and
ignore the heat extraction in the nucleation process. Really, the mathematical
structure of a diffusion equation is absolutely identical to the structure
of a heat transfer equation. So, all constructions for the condensation
heat extraction will be the same as for the substance consumption. It
will lead only for some renormalizations. That's
why only some corresponding remarks will be given. Any detailed results
can be found in \cite{book}.

We shall consider the situation of a metastable phase decay. It means
that in the initial moment of time all substance is in the vapor phase.
All heterogeneous centers are free from droplets.

\section{Profile  around the solitary droplet}

Due to the external influence in the initial moment of time one can observe
in  the system the homogeneous mother metastable phase with the particle
number density $n$ which is equal to some value $n_0$.
All heterogeneous centers are distributed rather homogeneously
in space with the number density $\eta_{tot}$.

The process of condensation can begin only when $n_0$ is greater than the
molecule number density $n_{\infty}$ in the vapor saturated over the plane
liquid. The power of metastability of vapor will be characterized by the
value of a supersaturation $\zeta$ defined as
$$
\zeta = \frac{n}{n_{\infty}} - 1
$$
The initial value of the supersaturation will be marked by $\zeta_0$.

Practically immediately around every center there will be formed an
equilibrium
embryo which has $\nu_e$ molecules of the condensing substance. The
value of $\nu_e$ is relatively small\footnote{In  comparison with the
characteristic number of molecules inside the droplet during the nucleation
period.} and there is no need to consider
the density profile around the equilibrium embryos\footnote{In fact it
will disappear rather fast. This leads to the slight variation of the
equilibrium embryo characteristics.  This variation will act on the gap.
But the final relaxation will be rather rapid.}.

The number of the equilibrium embryos in the unit volume\footnote{The
system of a unit volume will be considered.} will be equal
to $\eta_{tot}$. During the condensation process the number of the equilibrium
embryos $\eta$ will be decreased due to the exhaustion of the free
heterogeneous
centers
$$
\eta = \eta_{tot} - N
$$
where
$N$ is number of the supercritical embryos which will be called as the
droplets.

The last equation illustrates the main  specific feature
in kinetics of heterogeneous
condensation. Despite the simple form of the last relation the effect
is  very  complicate  because  $N$   depends  on  time  in  a  very
complicate
manner.

The effects of the density profile will be essential  also for $N$ and
one can not directly apply the results of \cite{PhysRevE}. One has to
determine the effect of the influence on $N$ even for the density profile
of the solitary droplet.

We shall call an approach where the law of embryos growth is found from
continuous model but there is no account of the profile around the droplets
as the "Additive approach" (AA).

Then  one can formulate  the evident

{\it Statement 1

The duration of the nucleation period\footnote{The period of nucleation
is the period of the relatively intensive formation of the droplets. It
can be proven that the end of this period is well defined due to the cut-off
of the intensity of the droplets formation.} and the characteristic sizes of
the droplets at the end of nucleation period are greater than those calculated
in AA}

Really, the existence of the density profile means that the part of substance
is going to be consumed from the regions where there is already no droplets
formation. This substance is consumed from the gap instead of unexhausted
regions as it is done in AA.

Then having repeated all constructions\footnote{In \cite{PhysRevE} the
AA was formulated for external conditions of dynamic type. For the situation
of decay the required hierarchical inequalities can be proven by the same
way. One can note that in \cite{PhysRevE} there is no special reference
on the type of condition when the required estimates are proven.}
 from \cite{PhysRevE} one can see\footnote{The barrier character of nucleation
is required here. It means that the magnitude of the activation barrier
height has the same order as the free energy of the homogeneous critical
embryos (may be it is $4$ or $3$ times smaller).}

{\it Statement 2

The characteristic size of the droplets at the end of the nucleation period
is many times greater than the size of the critical embryo. The main
role in vapor consumption is played by the supercritical embryos.}

and

{\it Statement 3

The characteristic time of the nucleation period duration is many times
greater than the time of relaxation to the stationary state in the
nearcritical
region. Then one can use the stationary rate of nucleation as the intensity
of the droplets formation in every current moment of time. }

Due to the statement 2 one has to investigate the profile around a growing
droplet. The problem is whether one has to consider the interference of
profiles around different droplets.

To solve this problem one has to use
the small parameter of the theory. Due to the statement 3 the rate of
nucleation
is equal to the stationary one. It can be taken from \cite{class}
$$
I_s = Z \eta exp(- \Delta F)
$$
where
$\Delta F$ is the height of activation barrier (taken in thermal units),
$\eta$ is the number of free (unoccupied by the supercritical embryos)
heterogeneous centers,
$Z$ is Zeldowitch factor.  Zeldowitch factor is the smooth function of
the supersaturation which is given by
$$
Z = \frac{W}{\pi^{1/2} \Delta \nu_e \Delta \nu_c}
$$
where
$W$ is kinetic factor,
$\Delta \nu_c$ is the halfwidht of the nearcritical region,
$\Delta \nu_e$ is the width of the equilibrium region.

Due to the rather small size of the critical embryo it is reasonable to
use the free molecular regime of the substance exchange\footnote{As long
as the characteristic size of the droplet during the nucleation is many
times greater that the critical size the diffusion regime of growth  for
the characteristic droplets is
quite reasonable.}. Namely, in this
regime an expression for the nucleation rate is well based. One has also
to note that the critical embryo is in equilibrium (unstable one) with
the metastable phase which implies no profiles of vapor density and regime
of the substance exchange is the free molecular one.

Under the free molecular regime $W$ can be calculated as
$$
W =  3 \frac{\zeta +1}{\tau} \nu_c^{2/3} \alpha
$$
where
$\nu_c$ is the number of molecules inside the critical embryo,
$\alpha$ is the condensation coefficient,
$$
\tau = 12 [(36 \pi v_l^2)^{1/3} n_{\infty} v_T ]^{-1}
$$
is the characteristic time,
$v_l$ is a volume per one molecule in a liquid phase,
$v_T$ is the mean thermal velocity of a molecule.

The value of $\Delta \nu_c$ is the halfwidth of the nearcritical region
and it can be calculated as
$$
\Delta \nu_c  = \sum_{\nu \leq (\nu_c + \nu_e) / 2 }
\exp(-F_c + F_{\nu}) \pi^{-1/2}
$$
where $\nu$ is the number of the molecules inside the embryo,
$F_{\nu}$ is the free energy of the embryo of $\nu$ molecules,
$F_c$ is the free energy of the critical embryo.
In continuous approximation it can be calculated as\footnote{Ordinary
$\Delta \nu_e$ is smaller than $\Delta \nu_c$ and an explicit summation
for $\Delta \nu_c$
is  quite reasonable.}
$$
\Delta \nu_c = |\frac{2}{\frac{\delta^2 F}{\delta \nu^2}} |_{\nu = \nu_c}
|^{1/2}
$$

The value of $\Delta \nu_e$ can be calculated as
$$
\Delta \nu_e = \sum_{\nu \leq (\nu_e + \nu_c) / 2 } \exp(-F_\nu + F_e)
$$
where
$F_e$ is the free energy of the equilibrium embryo.

Both $\Delta \nu_c$ and $\Delta \nu_e$ are rather smooth functions of
the supersaturation.

One can see that $I_s$ is a very sharp function of the supersaturation.
It means that the relatively small fall of the supersaturation leads to
the interruption of the droplets formation.

At least for all $\zeta > \ \zeta_0 / 2 $ one can show that
$ d^2 \zeta / d t^2 > 0$ and the is no long tail
of a size spectrum  with a small intensity.
It means that the interruption of the droplets formation leads
to the interruption of the nucleation process. So, the relative fall
of the supersaturation during the nucleation process is small.
One can come to the following statement

{\it Statement 4

During the nucleation period the relative variation of supersaturation
is small.
}

The last statement shows that there is no need to consider the interference
of profiles in order to change the rate of droplets growth.

On the base of mentioned expressions and the smallness of the relative
fall of supersaturation one can see the validity of the following
approximation
$$
I_s (\zeta) =           I_s (\zeta_0)
\exp  (\Delta F(\zeta_0) - \Delta F(\zeta))
$$
for the nucleation period.

Moreover one can linearize the height of the activation barrier over the
supersaturation and get
\begin{equation} \label{expap}
I_s (\zeta) =           I_s (\zeta_0)
\exp (
- \frac{d\Delta F(\zeta)}{d \zeta} |_{\zeta = \zeta_0}  (\zeta - \zeta_0)
)
\end{equation}
The validity of the last approximation depends on the concrete type of
heterogeneous centers but for the majority  of the heterogeneous centers types
the last approximation is valid.
For example, this validity can be directly proven for
ions.

One can explicitly calculate the derivative in the last
expression\footnote{Here
we assume the vapor to be an ideal gas and suppose the possibility to
present the free energy
of  critical and equilibrium embryos
 as an analytical  function  of  an  inverse
embryo
radius.}
$$
\frac{d \Delta F}{d \zeta } = -
\frac{1}{\zeta+1} (\nu_c - \nu_e)
$$

The smooth character of the last expression shows  the validity of
(\ref{expap})
one more time.

Then (\ref{expap}) can be rewritten as
\begin{equation} \label{expap1}
I_s (\zeta) = I_s (\zeta_0) \exp(\Gamma \frac{\zeta - \zeta_0}{\zeta_0}
)
\end{equation}
where
$$
\Gamma = - \zeta_0 \frac{d \Delta F}{d \zeta} |_{\zeta = \zeta_0} =
\frac{\zeta_0}{\zeta_0 + 1} (\nu_c (\zeta_0) - \nu_e (\zeta_0) )
$$

As far as the value of $\nu_c$ in going in the presented theory to infinity
the value of $\Gamma$ is also going to $\infty$. The real value of $\Gamma$
is very big. Certainly, one can consider the possibility of compensation
between $\nu_c$ and $\nu_e$ in the expression for $\Gamma$. Then one has
to mention thatdue to the barrier character of nucleation
 at least $\nu_c - \nu_e \geq \Delta \nu_c$. Having estimated
$\Delta \nu_c $ by the homogeneous value $\Delta \nu_c  \sim \nu_c^{2/3}$
one can see that $\Gamma \gg 1$ in any case.

The small value of $\Gamma^{-1}$ will be very important in further
constructions.

We see that essential dependence over the supersaturation occurs through
the height of activation barrier. It allows to give the interpretation
of the stationary rate of nucleation as the probability for the given
embryo to overcome the activation barrier. After the interpretation of
$I_s$ as the probability we can apply it to the arbitrary spatial point
of spatially unhomogeneous system. To use this interpretation
the natural requirement is the weak
unhomogenity of the system. Namely,
 the volume of the regions where
$$
\frac{ \zeta(r) - \zeta(r+ \sqrt{4 D t_s}  ) }{\zeta(r)} \ll \Gamma^{-1}
$$
is violated
has to be relatively small.

Here $D$ is the diffusion coefficient, $t_s$ is the time of relaxation
in the nearcritical region which can be estimated according to Zeldowitch
\cite{class}
$$
t_s \sim \frac{\Delta \nu_c^2}{W}
$$
One can use instead of $t_s$ the time $Z^{-1}$ which can be interpreted
as the mean time to overcome the nearcritical region.

Both these estimates are observed. Actually we need them only for those
regions where the intensity of the droplets formation isn't too small
in comparison with initial intensity. Certainly, the required property
is observed there.

Now we have to turn to the rate of embryos growth. According to the Statement
2 the characteristic size of droplets is rather big. Then it is more
reasonable
to use the diffusion regime of droplets growth.  Under the intermediate
Knudsen numbers one has to use an interpolation law  for the rate of embryos
growth (for example, see \cite{Fucs}). It will be important that all
expressions
for the embryos growth lead to the avalanche character of the substance
consumption.

The avalanche character of the substance consumption means that the quantity
of substance accumulated by  a  droplet strongly increases in
time. The most evident  manifestation of the avalanche consumption can
be seen in the free molecular regime of the substance consumption. The
most "weak" effect of the avalanche consumption can be seen under the
diffusion regime of the substance consumption.

The force of the iterations convergence in \cite{PhysRevE} is based on
this property. The property of avalanche consumption will be extremely
important in the further constructions also. That's why we take the diffusion
regime to have the worst situation and to grasp errors in all possible
cases.

In the diffusion regime of the vapor consumption the law of growth for
a  droplet (i.e. for the supercritical embryo) can be written in
the following way
$$
  \frac{d \nu}{dt} = \kappa \zeta \nu^{1/3}
$$
where
$$
\kappa = (\frac{2}{3})^{-1/3} 4 \pi n_{\infty} D (\frac{v_l}{2 \pi})^{1/3}
$$
is some constant.

The last expression is written in stationary approximation. The non-stationary
effects were investigated in many publications in details and they are rather
small ones.

One can see that the rate of the droplet growth is proportional to $\zeta$.
So, the rate of growth can be essentially changed only by the essential
relative variation of $\zeta$. Then according to Statement 4 one can see
that

{\it Statement 5

The rate of droplets growth during the nucleation period can be approximated
as a constant one.

}

The last statement is extremely  important  because  is  allows  to
analyze
the profile of the density initiated by a solitary droplet. Now this case
will be considered.

The approximately constant value of the supersaturation allows to integrate
the law of growth and get
$$
\nu (t) =
\gamma t^{3/2}
$$
where
$$
\gamma = (4 \pi)^{3/2}
(\frac{3 v_l}{4 \pi})^{1/2}
(\frac{ 2 \zeta  n_{\infty} D}{3} )^{3/2}
$$
and $t$ is the duration of the irreversible growth for the given droplet.

Consider the spherical system of coordinates with a center in the center
of droplet. Diffusion equation will be written as
$$
\frac{\partial n }{\partial t } = D \Delta n
$$
where $\Delta$ is the Laplace operator. Diffusion coefficient $D$ is
supposed to be approximately constant (there is a lot of a passive gas and
the density of a gas mixture is approximately constant).

The boundary conditions are the following
$$
n |_{r=\infty} = n(\infty)
$$
$$
n |_{r = R_d} = n_{\infty}
$$
where $R_d$ is the radius of a droplet. The values $n_{\infty}$ and
$n(\infty)$
are known parameters. The variable $r$ is the distance until the center
of the embryo.

The stationary approximation is suitable for the rate of droplets growth.
The errors are analyzed in \cite{Fucs} and they are small. But the stationary
solution can not  give any reasonable result for the density far from the droplet.
 Really, the
stationary solution is the following
\begin{equation} \label{st}
n(r) = n(\infty) - \frac{R_d}{r} (n(\infty) - n_{\infty})
\end{equation}
and has a very long tail. This tail leads to the infinite value of
$$
G = \int_0^{\infty} 4 \pi r^2  (n(\infty) - n(r)) dr
$$
which has to be the integral excess of the substance. This quantity has
to be in the droplet.  This contradiction shows that it
is absolutely impossible to use the stationary approximation for the density
profile around the droplet.  One has to introduce another approach.

One can see that if the first boundary condition will be changed by
$$
n |_{r=\infty} = n(\infty) (1- \Gamma^{-1} )
$$
then the rate of embryos growth will be changed unessentially. But the
level
$ n(\infty) (1- \Gamma^{-1} ) $ is the level when the nucleation stops.
So, one can see that during the nucleation period there is no interaction
between droplets through the change of the growth rate. Certainly, two
droplets can appear too close and act one upon another but the probability
of such coincidence is too small. That is why one can come to the "Principle
of a separate  growth of droplets during the nucleation period".

Now one has to prove that  at the distances $(5 \div 10) R_d$
from the droplet one can observe
the quasistationary profile.

One has to note that
\begin{equation} \label{vol}
v_l  / v_v \ll 1
\end{equation}
where $v_v$ is the partial molecule volume in a vapor phase. Really, the
last ratio is very small (for example, this ratio is $0.001$ for water
in normal thermodynamic conditions). But contrary to $\Gamma^{-1}$ one can not
consider it in all cases as zero.

Now one can introduce a formal parameter $l$  which attains big values
$$
l \gg 1 $$
but satisfies the following condition
\begin{equation} \label{relpr}
l^2 \frac{v_l}{v_v} \ll 1
\end{equation}
Due to (\ref{vol}) it is possible to do.

In the region $r \leq l R_d$ the stationary  profile is established after
$$
t_h = \frac{l^2 R_d^2}{4 D}
$$
It is necessary to show that
$$
s \equiv \frac{R_d(t+t_h) - R_d (t)}{R_d(t)} \ll 1
$$

Really,
$$
s \approx \frac{dR_d}{dt} \frac{t_h}{R_d}
$$
and
$$
s  \sim l^2 \frac{v_l}{v_v}
$$
which is a small value  according to (\ref{relpr}).
So, the stationary form of profile in the region $r<R_d l$ is proven.

As far as $\Gamma \gg 1$ and at least $\Gamma \gg l$ one can see that
in the region $r< l R_d$ there is no formation of new droplets. Then
this region isn't interesting for the theory and one can only observe
the region $r > l R_d$.

The previous notation is rather important. Namely this property allows
to use the  model with a point source. Really, one can consider only
the distances greater than $l R_d$. But at these distances the droplet can
be interpreted as a point.

Certainly, the point approximation of a droplet can not give an expression
for the rate of droplets growth because the boundary condition at $r =
R_d$ is absent. But the rate of growth is already known and can be used
directly as a known function of time. Really,
$$
\frac{d \nu}{dt} =
\lambda t^{1/2}
$$
where
$$
\lambda = 2^{5/2} \pi v_l^{1/2}   \zeta^{3/2} n_{\infty}^{3/2} D^{3/2}
$$

The action of a point source of a vapor consumption can be described in
a simple and suitable manner under the Green function formalism. The Green
function for the diffusion equation can be written in a following form
$$
Gr = \Theta(t) \frac{\exp(-\frac{r^2}{4Dt})}
{(4 \pi D t)^{3/2}}
$$

Then one can get the density profile by a simple integration
$$
n(r) =
n(\infty) -
\int_0^t
\frac{\lambda x^{1/2}}{(4 \pi D (t-x))^{3/2} }
\exp( - \frac{r^2}{4 D (t-x) } ) dx
$$

After evident transformations one can come to
\begin{equation} \label{argum}
\frac{\zeta_0 - \zeta}{\zeta_0} =
\sqrt{\frac{2}{\pi}} \sqrt{\frac{v_l}{v_v}} f(\beta)
\end{equation}
where
$$
\beta = \frac{r}{\sqrt{4Dt}}
$$
and
$$
f(\beta) = \int_{\beta}^{\infty}
(\frac{1}{\beta^2} - \frac{1}{x^2} )^{1/2} \exp(-x^2) dx
$$
It is important that the profile dependence on $t$ and $r$ is now going
through $\beta$.

Concrete form of $f(\beta)$ is drawn in fig. 1.
One can get for $f(\beta)$ an expression through  special functions
$$
f(\beta) = \frac{1}{2} \Gamma(\frac{3}{2}) \exp(-\beta^2)
\Psi(\frac{3}{2}, \frac{3}{2} ; \beta^2 )
$$
Here $\Gamma$ is the Gamma-function , $\Psi$ is the confluent hypergeometric
function.

One can get asymptotes for $f(\beta)$ at small and big values of $\beta$.
At small values
\begin{equation} \label{short}
f(\beta) \sim \frac{\sqrt{\pi}}{2} \frac{1}{\beta}
\end{equation}
which corresponds to the stationary solution (\ref{st}).

At big values of $\beta$ one can come to
\begin{equation} \label{long}
f(\beta) =
\exp(-\beta^2) \frac{1}{2 \beta^3}
\int_0^{\infty} x^{1/2} \exp(-x) dx \sim
\frac{\exp(-\beta^2)}{\beta^3}
\end{equation}
One can see that this asymptote radically differs from the stationary
solution. Namely this tail behavior ensures the convergence of the integral
for $G$. Certainly, the Green function formalism ensures a precise value
for $G$ which is introduced here as an external object.

Now one can turn to construct some approximation for the nucleation rate
around the growing droplet.

One can see that according to (\ref{expap1}) the behavior of the
supersaturation
is important when $\zeta_0 - \zeta \leq (2 \div 3) \zeta_0 / \Gamma $.
When $ \zeta_0 - \zeta \geq (2 \div 3) \zeta_0 / \Gamma$ the intensity
of the droplets formation is negligibly small.  From (\ref{expap}) one
can see that
$$
I_s (\zeta(r)) = I_s (\zeta_0 )
\exp(-\Gamma
\sqrt{\frac{2}{\pi}} \sqrt{\frac{v_l}{v_v}} f(\beta) )
$$
Then one can extract a positive parameter
$$
\sigma  \equiv
\Gamma^2 \frac{v_l}{v_v}
$$
which will be important in further constructions.

Due to $\Gamma \gg 1 $ one can easily see that
$$
\sigma \gg 1
$$
The last condition isn't necessary for further constructions, but it will
be rather important for manifestation of profile effects
in the nucleation process.

The last condition is also the most doubtful one because $v_l / v_v \ll
1$
and one has the ration of two big parameters with generally unknown result.
It is necessary to stress that condition $v_l / v_v \ll 1$ isn't so strong
as $\Gamma  \gg 1$. In frames of thermodynamic description $\Gamma \gg
1$ is the main condition necessary to give the thermodynamic description
  and $v_l / v_v \ll1 $ is the supplementary condition which  isn't
absolutely
necessary but slightly simplifies the theory.

In the situation of homogeneous condensation one has the "hidden contradiction
between the thermodynamic  description  and  the  relatively  intensive
nucleation".
Really, as far as in the homogeneous condensation $\Delta F = F_c \sim
\nu_c^{2/3}$ the limit $\nu \rightarrow \infty$ means $\Delta F \rightarrow
\infty$ and the rate of nucleation goes to zero. So, there is the
contradiction
between thermodynamic limit in the critical embryo description and the
observable\footnote{Which isn't too small.} rate of nucleation.

In the case of heterogeneous condensation there is no such contradiction
if there are some active centers of condensation.
Then the height of the activation barrier has no direct connection with
a number of molecules inside the critical embryo. For example, the halfwidth
of the nearcritical region estimated by a  homogeneous value has a value
$\sim \nu^{2/3}$ and goes to infinity when $\nu \rightarrow \infty$, but
the free energy decreases at the boundary of the nearcritical region
 only by one thermal unit. So, in certain sense
the case of heterogeneous condensation is more preferable for theoretical
description.

As a compensation for this advantage one has to note that the Statement
1 and Statement 2 are based on the homogeneous estimate for the activation
barrier height. These properties  can be violated. But as far as these
Statements are based on very strong inequalities one can accept their
validity.
The influence of nonstationary effects will be analyzed in a separate
publication
where the complete theory has been constructed.

Now one can analyze the profile of intensity of the droplets formation
around the already existing droplet. This nucleation rate profile is a
rather
sharp function which has a step-like behavior.

To show this we shall introduce two characteristic values of $\beta$
($\beta_{st}$ and $\beta_{fin}$) by relations
$$
f(\beta_{st}) =
\sqrt{\frac{\pi}{2}}
\sqrt{\frac{v_v}{v_l}}
\frac{\exp(-1/2)}{\Gamma}
$$
$$
f(\beta_{fin}) =
\sqrt{\frac{\pi}{2}}
\sqrt{\frac{v_v}{v_l}}
\frac{\exp(1/2)}{\Gamma }
$$

In the region $\beta > \beta_{st}$ the rate of nucleation
practically coincides with the unperturbed value $I_s (\zeta_0)$.
In the region $\zeta< \zeta_{fin}$ the rate of nucleation is negligible
in comparison
with the unperturbed value,  i.e. $I_s (\zeta(r)) \ll I_s (\zeta_0) $.

In some moment $t$ the values $\beta_{st}$, $\beta_{fin}$ initiate the
space distances $r_{st}$, $r_{fin}$ by the following expressions
$$
r_{st} = \beta_{st}  \sqrt{ 4 D t}
$$
$$
r_{fin} = \beta_{fin}  \sqrt{ 4 D t}
$$

For $\sigma \gg 1$ one can come to
$$ f(\beta_{st}) \ll 1 $$
$$ f(\beta_{fin}) \ll 1 $$
and
$$ \beta_{st} \gg 1 $$
$$ \beta_{fin} \gg 1 $$
Then one can use the asymptote (\ref{long}) and see that
$$
\frac{|\beta_{st} - \beta_{fin}|}{\beta_{st} + \beta_{fin}}
= \frac{1}{4\beta_{st,fin}}
\ll 1
$$
$$
\frac{|r_{st}-r_{fin}|}{r_{st} + r_{fin}} = \frac{1}{4 \beta_{st,fin}} \ll 1
$$

The real picture of nucleation is going in the time scale. In a fixed
space point $r$ one can introduce two characteristic times $t_{st}$ and
$t_{fin}$ by expressions
$$
t_{st} = \frac{r^2}{4 \beta^2_{st} D}
$$
$$
t_{fin} = \frac{r^2}{4 \beta^2_{fin} D}
$$

Before $t_{st}$ one can not observe any essential deviation of the nucleation
rate from the unperturbed value. After $t_{fin}$ the rate of nucleation
is very small.

One can get for the relative deviation
$$
\delta \equiv  \frac{t_{fin} - t_{st}}{t_{st,fin}}
$$
the following expression
$$
\delta \sim \frac{1}{ \beta_{st,fin}}
$$
So, the relative deviation is small.

Even in the situation of small $\sigma$ one can show with the help of
the asymptote (\ref{short}) that $\delta$ is rather small.

The step-like behavior\footnote{This step can take place even for $\beta
> \beta_{st}$ when $\sigma \ll 1$.}
 of intensity profile allows to introduce some characteristic
parameter $\beta_{eff}$ and consider the region\footnote{At $\sigma \ll
1$ the value  of $\beta_{eff}$ can be greater tan $\beta{st}$ and
$\beta_{fin}$.}
$$
\beta < \beta_{eff}
$$
as the exhausted region where  there is no nucleation more and in the
region
$$
\beta > \beta_{eff}
$$
the rate of nucleation is unperturbed\footnote{In all cases $\beta_{eff}
> \beta_{fin}$.}.

One has to choose $\beta_{eff}$ carefully. The problem is the possibility
of existence of a long tail of a density profile.
To grasp the situation of a small values of $\sigma$ one has to introduce
$\beta_{eff}$ in an integral manner.

One can introduce the excess of nucleation rate $\Delta I_s$ by the following
formula
$$
\Delta I_s =
I_s
\int_0^{\infty}
(1-\exp(-
\frac{\Gamma (\zeta_0 - \zeta(r) )}{ \zeta_0} ) )
4 \pi r^2 dr
$$
when $I_s$ is the unperturbed rate of nucleation.

On the base of the last expression one can get the excess of $N$ due to
the existence of the solitary profile. This value will be marked by $\Delta
N_{sol}$ and can be found as
$$
\Delta N_{sol} =
I_s
\int_0^t
\int_0^{\infty}
(1-\exp(-
\frac{\Gamma (\zeta_0 - \zeta(r) )}{ \zeta_0} )  )
4 \pi r^2 dr dt'
$$

Having used equation (\ref{argum}) one can come to
$$
\Delta I_s =
4 \pi      (4 D t)^{3/2}
I_s
\int_0^{\infty}
(1-\exp(-
\Gamma   \sqrt{\frac{2}{\pi}} \sqrt{\frac{v_l}{v_v}} f(\beta)  ) )
 \beta^2 d \beta
$$
Parameter $\Gamma \sqrt{\frac{2}{\pi}} \sqrt{\frac{v_l}{v_v}} $
has a constant value.

The value $\Delta N_{sol}$ can be now found as
$$
\Delta N_{sol} =
4 \pi      (4 D t)^{3/2}
I_s
\int_0^t
\int_0^{\infty}
(1-\exp(-
\Gamma   \sqrt{\frac{2}{\pi}} \sqrt{\frac{v_l}{v_v}} f(\beta)  ) )
 \beta^2 d \beta d t'
$$

The step-like approximation of the nucleation profile will lead to
$$
\Delta I_s^0 (\beta_{eff}) =
4 \pi      (4 D t)^{3/2}
I_s
\int_0^{\beta_{eff}}
 x^2 d x
$$

The value $\beta_{eff}$ has to be determined from
$$
\Delta I_s^0 (\beta_{eff}) = \Delta I_s
$$
Certainly, the value of $\beta_{eff}$ depends of     $
\Gamma   \sqrt{\frac{2}{\pi}} \sqrt{\frac{v_l}{v_v}}$.

The value of $\beta_{eff}$ initiates
$$
r_{eff} = 2 \beta_{eff} D^{1/2} t^{1/2}
$$

Inside the volume
$$
V_{eff} = \frac{4}{3} \pi r_{eff}^3
$$
there is no nucleation and outside this volume the rate of nucleation
is unperturbed.

Then one can imagine that around every droplet there exists the exhausted
region (ER) where there is no nucleation and the unexhausted region (UR)
where the nucleation remains unperturbed. The whole space now is divided
into two regions.

The volume $V_{eff}$ grows in time in the following way
$$
V_{eff} =
\frac{32}{3} \pi
\beta_{eff}^3  D^{3/2} t^{3/2}
$$
In a free molecular regime $V_{eff}$ will grow even faster.

For $b_{eff}$ one can get the simple expression
$$
\beta_{eff}^3 =
3
\int_0^{\infty}
(1-\exp(-
\Gamma   \sqrt{\frac{2}{\pi}} \sqrt{\frac{v_l}{v_v}} f(\beta)  ) )
 \beta^2 d \beta
$$
or
$$
\beta_{eff}^3 =
3
\int_0^{\infty}
(1-\exp(-
\sigma^{1/2}   \sqrt{\frac{2}{\pi}}  f(\beta)  )                  )
 \beta^2 d \beta
$$

For $\Delta N_{sol} $ one can obtain
$$
\Delta N_{sol} =  I_s (\zeta_0)  \int_0^t dt' V_{eff} =
\int_0^t dt' \frac{4}{3} \pi r_{eff}^3
$$

One can easy integrate the last expression and get
$$
\Delta N_{sol} = I_s (\zeta_0)
\frac{64}{15} \pi
\beta_{eff}^3  D^{3/2} t^{5/2}
$$
One can see that $\Delta N_{sol}$ is growing in time rather rapidly.
Namely this property illustrates the feature of the {\it "avalanche
consumption
during the first order phase transition"} in application to the
heterogeneous
nucleation.

For those situations where $\sigma \gg 1$ one can get
$$
\beta_{eff} \approx \beta_{st} \approx \beta_{fin}
$$
and
$
\beta_{eff}
$
satisfies the simple equation
$$
\exp(-\beta^2_{eff}) = \beta^3_{eff}  \sqrt{\frac{v_v}{v_l}}
\sqrt{\frac{\pi}{2}} \frac{1}{\Gamma}
$$
The last equation can be easily solved by iterations as far as $\beta_{eff}
\gg 1$ and $\exp(-\beta^2)$ is the most sharp function of the supersaturation.

Earlier when the principle of the separate growth was discussed then the
reason
of the absence of interaction between droplets was the low probability
to appear too close one to another only due to the smallness of such space
volume. Now one can see that the growing ER helps to
exclude the interference. Really, the essential deviation of the
supersaturation
from the ideal one can be seen in the region $r < R_d l$. It means
that the distance between the droplets with an interference must have
the order $2 R_d l$ Then the time distance between the moments of formation
of these droplets must be shorter than
$$
\Delta t_{init} \sim (\frac{R_d l }{\beta_{eff} D^{1/2} } )^2
$$
This time interval is many times shorter than the duration of the nucleation
period.

Every droplet forms rather rapidly after formation the ER of such a size
which guarantees that the rate of growth of the given droplet can not
be perturbed by a vapor consumption by the other droplets.

\section{Kinetic models of the global evolution }

Now one can construct the picture of nucleation in a whole system.
The main problem is to take into account the interference of the density
profiles. The interference through the rate of growth is absent, but there
is a simple overlapping of profiles which leads to the deviation of the
total nucleation rate over the volume from those calculated with account
of the additive
excesses around every droplet.

The overlapping of ER (even when this approximate formalism is used) is
very complex and can not be directly taken into account in precise manner.
Instead of writing the long expressions which can not be explicitly calculated
one can act in another manner: some simple approximate  models  for
kinetics
of the nucleation process will be formulated. From these models it will
be seen that they estimate the nucleation characteristics from below and
from above and lead practically to the similar results.
So, it will be shown that the
complex details of the ER overlapping has no strong influence on the real
characteristics of the phase transition.

At first one can consider the common feature of all models concerning
the exhaustion of free heterogeneous centers.

The rate of nucleation $I$ depends on time $t$ and on spatial point $r$
(the last behavior is the most complex). So, it is reasonable to consider
the mean (over the space) value of $I$ denoting it by $<I>$.  For $<I>$
one can write
the following expression
\begin{equation} \label{ttt}
<I> =
\frac{W_{free}}{W_{tot}} \frac{\eta}{\eta_{tot}} I_0
\end{equation}
where $I_0$ is the unperturbed rate of nucleation.
Here
$W_{free}$ is the volume of a region where the rate of nucleation is
unperturbed,
i.e. the UR of the whole system. The value $W_{tot}$ is the total volume
of the system (it equals to unity and is written only to clarify the
consideration).

Then as far as
$$
N = \int_0^t <I>(t') dt'
$$
one can get
$$
\eta = \eta_{tot} - \int_0^t <I> (t') dt'
$$

In the differential form the last relation can be written as
$$
\frac{d\eta}{dt} = -
<I>
$$
and with the help of (\ref{ttt}) it can be rewritten as
$$
\frac{d \eta}{dt} =  -
\frac{\eta}{\eta_{tot}}
\frac{W_{free}}{W_{tot}} I_0
$$

After the integration of the last expression one can come to
$$
\eta = \eta_{tot}
\exp(- \int_0^t
\frac{W_{free}(t')}{W_{tot}}
\frac{I_0}{\eta_{tot}} dt' )
$$

One has to note that heterogeneous centers aren't distributed homogeneously
with respect to the ER (or UR). Only free heterogeneous centers are
distributed
homogeneously with respect to ER. This fact has to be also taken into
account.

The problem is to determine the value of $W_{free}$. In different models
it will be determined in different forms.

{\it The model without overlapping.}

One can write
$$
W_{free} = W_{tot} - W_{exh}
$$
where
$W_{exh}$ is the volume where there is no further formation of the droplets.
Very approximately one can present it as the sum of ER around all appeared
droplets
$$
W_{exh} \approx \sum_i V_{eff}
$$
(the sum is taken over all already formed droplets). Certainly, the last
approximation
is rigorous only when there is no overlapping of the ER around different
droplets.

Having used an expression for $V_{eff}$ one can come to
\begin{equation} \label{oop}
W_{exh} = \int_0^t dt' <I>(t')
\frac{32}{3} \pi
\beta_{eff}^3  D^{3/2} (t-t')^{3/2}
\end{equation}

Having used an expression for $<I>$ one can come to the closed system
of nucleation kinetics equations
$$
W_{free} = W_{tot} -  \int_0^t dt'
\frac{\eta}{\eta_{tot}}
\frac{W_{free}}{W_{tot}} I_0
\frac{32}{3} \pi
\beta_{eff}^3  D^{3/2} (t-t')^{3/2}
$$
\begin{equation} \label{second}
\eta = \eta_{tot}
\exp(- \int_0^t
\frac{W_{free}(t')}{W_{tot}}
\frac{I_0}{\eta_{tot}} dt'               )
\end{equation}

In the quasihomogeneous limit (when there is no essential exhaustion of
centers) this system can be reduced to
$$
W_{free} = W_{tot} -  \int_0^t dt'
\frac{W_{free}}{W_{tot}} I_0
\frac{32}{3} \pi
\beta_{eff}^3  D^{3/2} (t-t')^{3/2}
$$
which can be written after the evident renormalization
$t \rightarrow  at, t' \rightarrow at'$ where
$a =  (I_0
\frac{32}{3} \pi
\beta_{eff}^3  D^{3/2})^{2/5}$  in the universal form
$$
W_{free} = 1 -  \int_0^t dt'
W_{free} (t-t')^{3/2}
$$

One has to note that in the general case the system of nucleation equations
can be solved with the help of methods developed in \cite{PhysRevE}.
At first one can solve the quasihomogeneous equation (it the Volterra
equation with a rather simple kernel which allows to apply the Laplace
transformation to solve it) and then on the base of quasihomogeneous equation
one can find the final rather precise expression using the second equation
as a formula for $\eta$.

Another variant is to solve numerically
the universal equation for $W_{free\ hom}$:
$$
W_{free\ hom} = 1 -  \int_0^t dt'
W_{free\ hom} (t-t')^{3/2}
$$
Then one has the universal function $W_{free\ hom}$. Then
one can find $\eta$ as
$$
\eta = \eta_{tot}
\exp(- a ^{-1} \int_0^t
\frac{W_{free\ hom}(t')}{W_{tot}}
\frac{I_0}{\eta_{tot}} dt' )
$$

The last expression leads to the formula for $<I>$ as:
$$
<I> =
\frac{W_{free\ hom}}{W_{tot}}
\exp(- a^{-1} \int_0^t
\frac{W_{free\ hom}(t')}{W_{tot}}
\frac{I_0}{\eta_{tot}} dt'
  I_0                       )
$$
The justification of such approach is analogous to \cite{PhysRevE}.
The physical reason is very simple: when there is no exhaustion of
heterogeneous
centers then the solution is found precisely, when there is an essential
exhaustion of centers there is no need to know $W_{free}$ with a high
precision because the converging force of the operator of the right
hand of the second equation in the nucleation equations system is extremely
high.

Now we shall take into account the overlapping. It can be done rather
approximately.

{\it The model with chaotic overlapping }

The matter of discussion is the correct expression for $W_{free}$ which
can not be got absolutely precisely. Now the more reasonable expression
for $W_{free}$ will be presented. Certainly, this will lead to a more
complicate equation which will be more difficult to solve.

One can use the differential approach to write the expression for $W_{free}$.
Having written the evident relation
$$
\frac{d W_{free}}{dt} =
 -
\frac{d W_{exh}}{dt}
$$
one has to invent an approximation for $d W_{exh} / dt$. Here the following
approximation
$$
\frac{d W_{exh}}{dt} \approx \frac{ d \sum_i V_{eff} }{dt}
\frac{W_{free}}{W_{tot}}
$$
(the sum is taken over all droplets) will be used.
It corresponds to the following approach:
{\it The probability of the absence of overlapping of the
new parts of ER around the given droplet with other ER
 is proportional to the free volume
of the system.
}

The last supposition seems to be rather reasonable.

The value $d \sum_i V_{eff} / dt$ can be rewritten as
$$
\frac{d \sum_i V_{eff}}{dt} =
\sum_i \frac{d V_{eff}}{dt}
$$
The last value can be easily expressed through $<I>$ as
\begin{equation} \label{out}
\sum_i \frac{d V_{eff}}{dt} =
\frac{3}{2}
 \int_0^t dt'
 <I>(t')
\frac{32}{3} \pi
\beta_{eff}^3  D^{3/2} (t-t')^{1/2}
\end{equation}
due to (\ref{oop}).

Then
$$
\frac{dW_{exh}}{dt} =
\frac{3}{2}
 \int_0^t
 <I>(t')
\frac{32}{3} \pi
\beta_{eff}^3  D^{3/2} (t-t')^{1/2}
dt'
W_{free}(t)
$$
and
$$
\frac{dW_{free}}{dt} = -
\frac{3}{2}
 \int_0^t
 <I>(t')
\frac{32}{3} \pi
\beta_{eff}^3  D^{3/2} (t-t')^{1/2}
dt'
W_{free}(t)
$$
Having used an expression for $<I>$ one can come to
$$
\frac{dW_{free}}{dt} = -
\frac{3}{2}
 \int_0^t
\frac{W_{free}}{W_{tot}} I_0
\frac{\eta}{\eta_{tot}}
\frac{32}{3} \pi
\beta_{eff}^3  D^{3/2} (t-t')^{1/2}
dt'
W_{free}(t)
$$

Together with (\ref{second}) the last equation forms the closed system
of the nucleation equations in the second model.

The previous equation  can be integrated which gives
$$
\ln W_{free}= -
 \int_0^t
\frac{W_{free}(t')}{W_{tot}} I_0
\frac{\eta(t')}{\eta_{tot}}
\frac{32}{3} \pi
\beta_{eff}^3  D^{3/2} (t-t')^{3/2}
dt'
+ const
$$

Due to the initial conditions the $const$ is equal to zero. Having introduced
the function
$ F = - \ln  W_{free} $ one can get for $F$ the following system of equations
$$
F (t) =
 \int_0^t
\exp(-F(t')) I_0
\frac{\eta (t') }{\eta_{tot}}
\frac{32}{3} \pi
\beta_{eff}^3  D^{3/2} (t-t')^{3/2}
dt'
$$
$$
\eta = \eta_{tot}
\exp(- \int_0^t
\exp(-F(t'))
\frac{I_0}{\eta_{tot}} dt'                               )
$$

One can see that the system of condensation equations is identical to
the system of nucleation equations in AA.
It was completely analyzed in \cite{PhysRevE}.
Certainly, the parameters in the system will be another.

The last system can be rewritten after the evident renormalization as
$$
F(t) = \int_0^t \exp(-F(t'))(t-t')^{3/2} \theta(t')  dt'
\equiv \hat{F}(F,\theta)
$$
$$
\theta (t) = exp[-A \int_0^t \exp(-F(t')) dt' ]
\equiv \hat{\theta} (F)
$$
where
$\theta (t) = \eta(t) / \eta_{tot} $ and $A$ is some known parameter.

This system can be solved by iterations defined as
$$
F_{i+1} = \hat{F} (F_i, \theta_i)
$$
$$
\theta_{i+1} = \hat{\theta} (F_i)
$$
with
$F_0 = 0$, $\theta_0 = 1$.

For $F_i$, $\theta_i$ one can get the chains of inequalities
$$
F_0 < F_2  .. < F_{2i}< ... < F< ... < F_{2i+1} < ... < F_3 < F_1
$$
$$
\theta_1 < \theta_3 < ... < \theta_{2i+1} < ... < \theta < ... < \theta_{2i}
< ... < \theta_2< \theta_0
$$
Then one can estimate errors in $F_i$, $\theta_i$.

One can use also another methods analogous to those described in
\cite{PhysRevE}.

The similarity of the the condensation equations in AA and in the second
model is extremely important for the transition towards collective character
of vapor consumption which is analyzed in \cite{PhysRevE}.

The physical ground of the considered model is the chaotic overlapping
of ER. Namely, the chaotic overlapping lies in the base of approximation
used here. But due to the spherical form of every ER the overlapping isn't
absolutely chaotic.

What can be done in such a situation? In the next model we shall show
that concrete type of overlapping isn't very important.

To finish with the second  model we shall show the same method of its solution
as that of the first  model.

One can also formulate the quasihomogeneous equation as the following
one
$$
F_{hom} (t) =
 \int_0^t
\exp(-F_{hom}(t')) I_0
\frac{32}{3} \pi
\beta_{eff}^3  D^{3/2} (t-t')^{3/2}
dt'
$$

Then $\eta$ can be approximately found as
$$
\eta = \eta_{tot}
\exp(- \int_0^t
\exp(-F_{hom}(t'))
\frac{I_0}{\eta_{tot}} dt'
$$

The quasihomogeneous equation can be renormalized. After renormalization
$z \rightarrow at$, $t' \rightarrow a t'$ where
$$
 a = (
 I_0
\frac{32}{3} \pi
\beta_{eff}^3  D^{3/2}
)^{2/5}
$$
one can  transform the quasihomogeneous equation into the universal form
$$
\ln W_{free\ hom} (t) = -
 \int_0^t
W_{free\ hom} (t')
 (t-t')^{3/2}
dt'
$$

{\it  The model with the formation of droplets inside the ER }

The third model will show that the role of overlapping isn't so essential
as it can be imagined from the first point of view.

Suppose that the new droplets can appear also in the ER of the already
existed droplets.
Then instead of (\ref{out}) one has to use
$$
\sum_i \frac{d V_{eff}}{dt} =
\frac{3}{2}
 \int_0^t dt'
 I_0 \frac{\eta(t')}{\eta_{tot}}
\frac{32}{3} \pi
\beta_{eff}^3  D^{3/2} (t-t')^{1/2}
$$

Then
$$
\frac{dW_{exh}}{dt} =
\frac{3}{2}
 \int_0^t
 \frac{\eta(t')}{\eta_{tot}} I_0
\frac{32}{3} \pi
\beta_{eff}^3  D^{3/2} (t-t')^{1/2}
dt'
W_{free}(t)
$$
and
$$
\frac{dW_{free}}{dt} = -
\frac{3}{2}
 \int_0^t
 I_0 \frac{\eta(t')}{\eta_{tot}}
\frac{32}{3} \pi
\beta_{eff}^3  D^{3/2} (t-t')^{1/2}
dt'
W_{free}(t)
$$
Together with (\ref{second}) the last equation forms the closed system
of the nucleation equations in the third  model.

The substance balance equation of the system can be integrated which gives
$$
\ln W_{free}= -
 \int_0^t
 I_0
\frac{\eta(t')}{\eta_{tot}}
\frac{32}{3} \pi
\beta_{eff}^3  D^{3/2} (t-t')^{3/2}
dt'
+ const
$$

Due to the initial conditions the $const$ is equal to zero. Having introduced
the function
$ F = - \ln  W_{free} $ one can get for $F$ the following system of equations
$$
F (t) =
 \int_0^t
 I_0
\frac{\eta (t') }{\eta_{tot}}
\frac{32}{3} \pi
\beta_{eff}^3  D^{3/2} (t-t')^{3/2}
dt'
$$
$$
\eta = \eta_{tot}
\exp(- \int_0^t
\exp(-F(t'))
\frac{I_0}{\eta_{tot}} dt' )
$$

This system corresponds to the first iteration in the iteration solution
of the second model by the method of  iterations of a type "a" described
in \cite{PhysRevE}.
It is also
described here. But for $\eta$ the whole set of iterations has been taken
(see details in \cite{PhysRevE}).

One can slightly  modify the model and suppose that in the expression
for $\eta$ one can use the same approximation for $<I>$  as in equation
for $W_{free}$.  Then the system of equation will precisely correspond
to the first iteration in the iteration solution in \cite{PhysRevE}.

Actually one needn't to analyze these models in details following
\cite{PhysRevE}
but has  only to note that all these solutions are practically similar. One
can see the solution of the third model (both variants) after the same
renormalizations as in the second model.

Now one has to explain why the third model is rather accurate. One can
do it only with the help of results obtained in \cite{book}.

In \cite{book} it was noted that when the power of the kernel
$(t-t')$ is rather big then the solution of the quasihomogeneous equation
weakly depends on concrete value of a power.

Another important notation shows that when the power of $(t-t')$ is
extremely high
then the ER
of the given (first) droplet formed inside the ER of another
(second) droplet practically can not go outside
the ER of that (second) droplet.
Then the third model  is absolute adequate in this situation.
The same feature can be seen directly from the iteration procedure results.

Having combined these two notations one can see that on one hand the second
model
is close to the situation with great  power of the kernel and in the situation
with the big kernel the third model is suitable.

Now it is possible to explain why the overlapping isn't so important as
it can be imagined. Really, as far as the power of kernel is big and
one can observe the avalanche consumption one can see the following
qualitative
picture

\begin{itemize}

\item
Practically during all period of nucleation the total  ER is small and there
is
no problem of overlapping.

\item
At the end of the nucleation period the  total ER is going to occupy the
essential
part of the volume and
a few moments late it occupies  all the volume of the system.
This process is very rapid. It stops
the nucleation.

\end{itemize}

This picture shows that there is no strong influence of the overlapping
on the nucleation process (except the cut-off).

As the result the nucleation description is completed. One can use both
the
second and the third model to get the nucleation description. How to solve
these equations is also described here.

{\it   The method of universal solution }

The main idea of the theory presented in \cite{PhysRevE} was to consider
the quasihomogeneous equation, to get the universal solution and then on the
base of this solution to calculate the number of free heterogeneous centers.
As the result one can get an expression for $<I>$ and can calculate the
total number of droplets appeared on the nucleation process.

Here one can follow the same idea. But one can make this idea more deep.
One needn't to formulate the universal quasihomogeneous equation. Instead
of equation one can formulate the universal model.

The model will be the following

\begin{itemize}

\item
-the rate of nucleation $<I>$ can be found from
$$
<I> = I_0 \frac{W_{free}}{W_{tot}}
\frac{\eta}{\eta_{tot}}
$$

\item
- with intensity $I_0$ the droplet appears in the arbitrary point of the
system which belongs to the free volume.

\item
-the value $W_{free}$ can be found by exclusion of all ER around the already
existing droplets.

\item
- if the point is occupied by the ER of any droplet then the droplet can
not be formed.

\item
- the size $r_{eff}$ of ER grows in time according to
$$
r_{eff} = 2 \beta_{eff} D^{1/2} t^{1/2}
$$

\item
-the initial condition is the absence of droplets and random distribution
of centers.

\end{itemize}

With the  proper renormalization of time $t$ and size $r$ one can cancel all
coefficients. Then this process will be the universal one and as the result
$W_{free}$ is the universal function of time.
Then one can directly apply (\ref{second}) and get the number of free
heterogeneous centers (after the proper renormalization)\footnote{This
is quasihomogeneous approach method}.

The modification for the dynamic conditions requires to put instead of
$I_0$ the value $I_0 exp(ct)$ with some parameter $c$
determined by external conditions  and to change the
lower limit $0$ of integration to $- \infty$ \cite{PhysRevE}.

The main constructions of
 the theory will be absolutely the same but the forms of characteristic
curves changes radically.

        \section{Numerical results}

Numerical simulation in the nucleation kinetics plays at least
two important roles. The first role is the standard comparison with
the approximate models to observe their quality. The second one is
more specific and it is concerned with some universal dependencies
in the nucleation kinetics.

In the AA to the nucleation kinetics it was shown that the adequate approach
can be presented on the base of the quasihomogeneous  solution
\cite{PhysRevE}.
Despite the dynamic conditions considered in \cite{PhysRevE} one can observe
this property in the situation of the metastable phase decay also. Recall
the reasons for such approach.

The formal reason is the careful analysis of the iteration procedure proposed
in \cite{PhysRevE}. Really, the final result for the total number of droplets
appeared in the nucleation process is given by the second iteration (see
 iterations of type "a" in \cite{PhysRevE}) for
the relative number of free heterogeneous centers. This
iteration is based only on the first iteration for the supersaturation.
The value of supersaturation  there
 is calculated without any account of the heterogeneous centers
exhaustion. So, one can see that the final result can be obtained on the
base of the supersaturation in the quasihomogeneous approximation. This
approximation
can be  more sophisticated than the first iteration. Namely this approach
was used in \cite{PhysRevE} where the precise
quasihomogeneous universal solution was chosen
as base          for the final results.

The physical reason of such behavior is rather simple. The main role in
the vapor consumption belongs to the droplets of relatively large
sizes. We have already
marked this fact. But moreover, due to the avalanche character of the vapor
consumption the main role in vapor consumption is played by the relatively
large droplets which are formed in the first moments of time of the nucleation
period. When the effect of the centers exhaustion is essential already
in the first moments of nucleation period\footnote{More precisely one can
define these "first moments of time"
as $2/5$ of the nucleation period duration
(under the free molecular regime it is $1/4$ of the nucleation period
duration).
The reason for concrete value can come from the iteration procedure.}
then at the end of nucleation period already all centers will
be exhausted. Then the result is evident - all centers are the centers
of droplets.  Due to the
high force of convergence in this situation
this result can be gotten under the rather arbitrary behavior
of supersaturation (even including the quasihomogeneous case).
In the opposite case the exhaustion of heterogeneous centers during the
first moments of the nucleation period isn't essential and one can get
the quasihomogeneous behavior of supersaturation.

This effect will be called as the "approximate separation of the
heterogeneous
and homogeneous problems". It is based only on the avalanche consumption
of the metastable phase.
So, there are no objections to see this effect also in the situation with
the density profiles. Then it is extremely important to get solution in the
quasihomogeneous situation and clarify whether it can be presented in
the universal form.

The universal form of the quasihomogeneous solution can be easily seen
in the situation of profiles also. In the AA there was no specific space
scale because the consumption took place homogeneously from all space
points of the system. Here in the situation of profiles
there is the elementary space scale and one can
choose the space scale as to have that the linear size of ER around the
droplet
is growing\footnote{Certainly, the power has to be conserved.}
as $t^{1/2}$. The time scale has to be chosen as to get that in the free
volume equal  to the total volume of the system  one can see  the appearance
of
one droplet in the unit of time. As long as the functional dependencies
of the nucleation rate and the radius of ER are not identical one can do such
a renormalization without any problems. So, we see that the pseudohomogeneous
case allows the universal description.

The process of the heterogeneous centers exhaustion destroys this universality
and one has to act as in\footnote{Here we use  slightly more simple way.}
  \cite{PhysRevE}. The total number of droplets has to be
approximately calculated as
\begin{equation} \label {ror}
N_{total} =
\eta_{tot} [ 1- \exp(- \frac{N_{hom}}{\eta_{tot}} )]
\end{equation}
where
$
N_{hom}$ is the number of the droplets appeared in the quasihomogeneous
situation (under the same parameters).
This formula can be also used for all approximate models described earlier.

For the numerical simulation it was convenient to consider the cube box with
a size $10$. The rate of the ER growth is chosen as
$$ \frac{dR}{dt} = 100 t^{1/2}
$$
where $R$ is the radius of ER.
The rate of nucleation is chosen as to have one attempt of a
new droplet formation in the system
during $dt = 0.002$. This attempt is governed by the random procedure.
It may lead to the point in one of the ER and then no droplet will be
formed. In the opposite situation when the point is indicated out of all
ER of the already existing droplets there will be a new formation of a
droplet.

One has to stress that the random procedure has one specific negative
feature.
In the standard numerical procedures
the next random coordinate is calculated on the base of the previous ones.
So, if the current coordinate lies near the center of the already existing
ER then the next coordinate will be also near the center of another ER.
These correlations lead to the necessity to consider the large system.
In the system under consideration the number of droplets appeared in the
quasihomogeneous situation will be near $500$.  Nevertheless the square
mean fluctuation will be about $20$. The careful consideration can show
that the error induced by the substitution of the periodic boundary conditions
by the zero ones has the same power as the mean square error.
It can be directly seen by the numerical simulation.  It is explained
by the evident notation that the characteristic overlapping of profiles
is about the mean profile size. We shall call this feature as the property
of "moderate overlapping".  This fact can be proven analytically.
One can also analytically prove that the
effect of fluctuations is really caused by the mentioned phenomena.

The mean value of the total droplets number is equal to $504.8$ (under
the zero boundary conditions). This value has to be put into the previous
formula.

The avalanche character of the vapor consumption is illustrated by fig.
2-4. Three different moments of time $t=0.5$, $t=1$ and $t=1.5$ are
chosen as characteristic values.  The cross section is drawn.
The dashed regions correspond to the
ER of the already existing droplets. The black regions correspond to the
overlapping\footnote{Only when the distance between centers is approximately
the odd
one.} of ER.

The number of heterogeneous centers in such a system can be arbitrary.
Certainly the effects of the centers exhaustion will be important when
the number of centers is small (in comparison with $N_{hom} = 504.8$).
The pictures for $\eta_{tot} = 50$ are drawn
in fig. 5-7 for $t=1.5$, $t=3$, $t=6$.
One can see that the number of ER is smaller than in the quasihomogeneous
case.  The size of ER when the free volume
is going to  be exhausted  is larger than
the size of ER in the quasihomogeneous case. The time necessary to cover
all volume by ER  is greater than in the quasihomogeneous case. It doesn't
mean that the duration of the nucleation period will be longer (simply
all centers will be exhausted and this means the end of nucleation).
Moreover, the duration of the nucleation period in the situation of the
relatively
small number of heterogeneous centers will be shorter than in the
quasihomogeneous
case.

One can also see that the avalanche character of the vapor consumption
here will be  more smooth than in the quasihomogeneous case. Certainly, in
the quasihomogeneous case the appearance of the new ER helps to consume
the vapor phase in the avalanche manner. But in the situation with the small
number of centers there is no need to consider the process carefully because
the exhaustion of centers leads to the evident result of condensation
- the number of the droplets equals to the number of centers.

Now it is evident that the main object of our interest will be the
quasihomogeneous
case. The relative rate of nucleation in this case is shown in fig. 8.
Here the rate of nucleation is averaged over $100 dt \equiv 0.2$ and over
$16$ attempts. So, the rate of nucleation is rather smooth function.

The relative rate of nucleation is compared in fig.8 with the mentioned
models. Certainly, the rate of nucleation defines the spectrum of sizes
when we take as the size of embryo those characteristic which has the rate
of growth independent on the  value  of  size.  For  the  diffusion
regime this
characteristic is the number of molecules in the power $2/3$.

One can see in fig. 8 three different curves and some solitary points.
The solitary points correspond to the numerical simulation of the
quasihomogeneous
case  and  three  curves  correspond  to  three   models   in   the
quasihomogeneous
case.

The  shortest spectrum is in the first model. The careful look shows
 that this line is doubled. This occurs because there the ideal variant
of the first model is also drawn. This ideal variant corresponds to
$W_{free} / W_{total} \equiv 1$ in the subintegral function. The coincidence
means that the main role in the first model is played by the relatively
big droplets which were formed at $W_{free} = W_{total}$.

The  longest spectrum corresponds to the second model. This curve is
very close to the intermediate curve which corresponds to the third model.
The approximate coincidence  of the second and the third models shows that
both of them are valid and the role of the relatively big droplets here
is the main one.

One also see that the first model isn't too far from the real solution.
This allows to present the rigorous estimates for the nucleation rate.
Certainly the first model is the estimate of the real process from below.
It gives the number of the droplets about $20$ percents less than the
numerical simulation.

An estimate for the nucleation rate from above can be gotten by the following
way. From the first model it follows that until $t=0.52$ (practically this case
is
drawn in fig. 2) the rate of nucleation is near the ideal value and
the deviation is less than $15$ percents. So, one can say that the period
$0<t<0.52$ corresponds to the absence of overlapping (the first model
is the estimate from above). Then one can consider the process where the
total volume is exhausted only by the ER of the droplets appeared at
$0<t<0.52$
in a random manner.
The distribution of the centers of ER of such droplets is evidently random.
This model certainly gives the estimate from above for the nucleation process.
The total number of the droplets is only $15$ percents greater than the
result of the numerical simulation. As a conclusion one can state that
two suitable estimates
from below and above are obtained.

The proximity of the last estimate to the real solution justifies the
supposition that the main role in vapor consumption belongs to the droplets
of the relatively big sizes appeared in the system practically free from
the ER. This supposition can be also justified in the analytical manner.

One can see that the second and the third models are rather close to the
real solution but don't coincide with it. There are at least two reasons
of such deviation. The first one is the presence of the strong correlations
in the real system - if two ER overlap in some moment of time then the
power of overlapping can only grow in time. It hasn't  a random
character as it is stated
in the second and third models.

This effect can be taken into account in a rather simple manner. It is
sufficient to
consider two spheres and calculate the power of overlapping as the function
of the distance and time (it is the simple geometrical problem).
Unfortunately the answer can be written only in a very complicate form.
If we have two ER with radii $R_1$ and $R_2$ correspondingly at a distance
$l$ between their centers  and $l > max(R_1, R_2)$ then the volume of overlapping is
$$
V_{over} =
\frac{2 \pi R_1^3}{3} (1- 2 cos \varphi_1 + cos^3 \varphi_1)
+
\frac{2 \pi R_2^3}{3} (1- 2 cos \varphi_2 + cos^3 \varphi_2)
$$
where
$$
cos
\varphi_1 = \frac{-R_2^2 + R_1^2 + l^2}{2 R_1 l }
$$
$$
cos
\varphi_2 = \frac{-R_1^2 + R_2^2 + l^2}{2 R_2 l }
$$
Certainly, this result can not lead to the compact form of the balance
equation. It will be difficult to solve it analytically.

The second reason is the "moderate overlapping" problem. This property means
that
actually there is an interaction through overlapping in ensemble of
several droplets. Earlier this property was extracted \cite{book2} in a case
of a special effective length of ER. Now we see that this property has
a
rather general sense. The way to solve this problem proposed in \cite{book2}
is very complicate and leads to some uncertain relations.

How one can overcome all these problems? Actually one has no need to do
it. The simple numerical simulation takes into account all these effects
and gives the universal solution. Really we need only  one number -
the total number of appeared droplets. It is known. Then one can forget
about all mentioned difficulties.

Now one can analyze the heterogeneous case explicitly. The suitable
approximation
is given by (\ref{ror}). One has to substitute instead of $N_{hom}$ the
number of droplets given by the corresponding model.

The relative error of approximation (\ref{ror}) is drawn in fig. 9 for
the first model, in fig. 10 for the second model, in fig. 11 for the third
model. It is rather small for all models.
 For the third model it is practically negligible.
It is clear because the third model is based on the free volume approximation.
Then
the source of error can be found only in exhaustion of centers without any
influence
on the growing of ER. Without this influence the power of exhaustion can
not essentially act on the process of nucleation.

One can fulfill the same analysis for the numerical simulation. In fig. 12
the relative error of (\ref{ror}) for numerical simulation is drawn. Here
in (\ref{ror}) the value $N_{hom} = 504.8$ from the numerical simulation is
used.
The  result  is  compared   with   the   computer   simulation   of
heterogeneous condensation.
This simulation is rather simple. One can take the procedure for the
quasihomogeneous case
but put the center of the new droplet with probability $\eta/ \eta_{tot}$.
Every time when this point will be out of ER we shall reduce $\eta$ as
$\eta \rightarrow \eta - 1 $.

One can see that the
relative error is very small. We don't use the averaging over
(this is the reason why there is no smooth curve)
many attempts in order to see that the error of (\ref{ror})
 has the scale of the mean
square error of numerical simulation\footnote{For the system with $500$
droplets.}.
So, there is no need to use the more sophisticated approach.

The solution of the problem now is completed.

The generalization for the conditions of dynamic type is absolutely analogous
to \cite{PhysA}.
The convergence due to avalanche consumption is more weak and one has
to use instead of approximation (\ref{ror}) more sophisticated procedure
described in \cite{PhysRevE}. Certainly, the universal constants have
to be calculated by the numerical simulation with profiles.

The generalization for the arbitrary regime of the droplets growth can
be done absolutely analogous to \cite{PhysA}. This generalization is
based on the similarity of the functional forms obtained here and in the
AA. This similarity lies in the base of universality formulated in
\cite{book},
\cite{book2}.

One can see that the theory of condensation with profiles taken into account
presents the picture which is absolutely  different  from  the  AA.
Nevertheless
in many situations the result of experiment coincides with the result
of the AA. One has to explain this coincidence. It is rather formal one.

Really, in any experiment it is more convenient to have a small
system and to get many droplets. The rate of nucleation has to be taken as
a rather high one.
 So, the supersaturation is relatively high
and parameter $\sigma^{-1}$ isn't a real small parameter of a theory\footnote{
It isn't necessary for consideration here, but it has to be small due
to the possibility of thermodynamic description of the critical embryo.}.
 Then as it is
shown in \cite{book2} the AA gives the write qualitative
result despite the wrong base of consideration. The reason lies in the
fact that at small $\sigma$ the main quantity of a substance is lying in
the tail of profile. The tail of profile is rather thin and
can be  be taken into account by the AA. In   \cite{book2}
the correction term for the AA at small $\sigma$ is also presented.

The last important feature to mention is the movement of the embryos
boundaries.
This problem is widely discussed in the determination of the rate of regular
growth for the supercritical embryos. In different systems the effect
of influence of the
boundary movement on the rate of growth is different. We have to
note that in  theory presented here
 the rate of the  embryos growth is an external value
which is assumed to be known\footnote{It is really known practically for
all systems.}.

Another problem is the adequate account of the effect of boundary movement
in construction of the ER. One can see that in the first part of publication
it is already shown that the effect of the boundary movement is small. Here
we shall present the abstract  arguments for such conclusion.

To use the thermodynamic approach the initial power of the mother phase
metastability has to be relatively small.
Together with the Maxwell's rule it leads to the following final result
of the phase transition:
"Only relatively small part of the system volume is occupied by the new
phase."
It isn't in contradiction with the property that all volume is occupied
by ER.
So, the final state of the system is the practically saturated mother phase
in practically all volume of the system and the small volume (distributed
over all system) occupied by a new phase.

As the result one can see that the process of a substance consumption
(extraction) leads to the saturation in a  volume relatively large in
comparison with a volume of the new phase embryo.

The mother phase can not be undersaturated (then there will be  the
disappearance of embryos). So, as long as even the saturated phase has
to be spread on large distances $l_0$ one can state that the growth of embryo
produces the perturbation over the large distances $l > l_0$.
To have an interruption
(relative interruption in comparison with an ideal nucleation rate)
of a new phase formation one needs a very small reduction of the power
of metastability\footnote{Relative reduction has to be $\Gamma^{-1}$ where
$\Gamma \gg 1$ has the scale of the number of molecules in the critical
embryo.}. So, this reduction is attained at the distances $l' > l$ which
are very large in comparison with the embryos linear size. Then one can
use the point source approximation as it was done in the first part and
forget about the boundaries movement. The negligible character of the boundary
movement is proven now for all possible systems.

The heat extraction and account of all other intensive parameters of
description
can be done analogously to \cite{book}.

\pagebreak



{\it
\begin{center}
Fig. 1

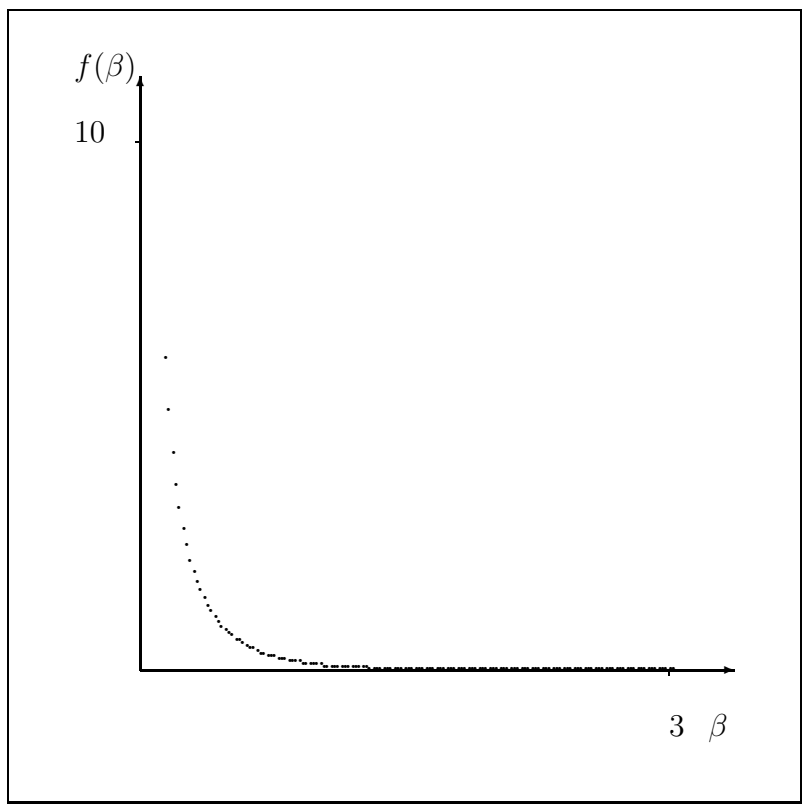
The form of $f(\beta)$.

\end{center}

}

\pagebreak

%

{\it
\begin{center}
Fig. 2

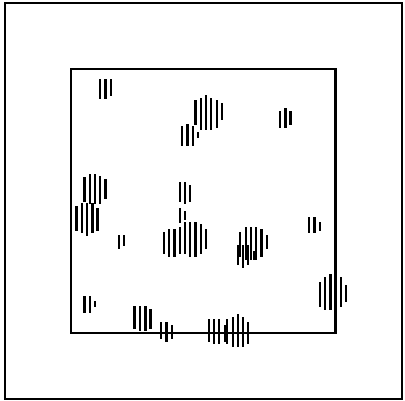
The cross section of the system at $t=0.5$ for the quasihomogeneous situation.

\end{center}

}

\pagebreak

%

{\it
\begin{center}
Fig. 3

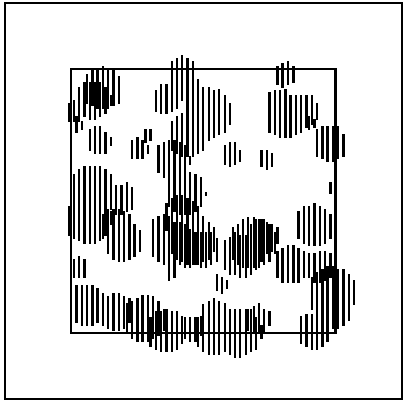
The cross section of the system at $t=1$ for the quasihomogeneous situation.

\end{center}

}

\pagebreak

%

{\it
\begin{center}
Fig. 4

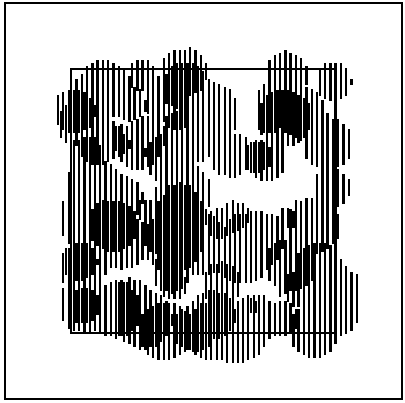
The cross section  of the system at $t=1.5$ for the quasihomogeneous
system.

\end{center}

}

\pagebreak

%

{\it
\begin{center}
Fig. 5

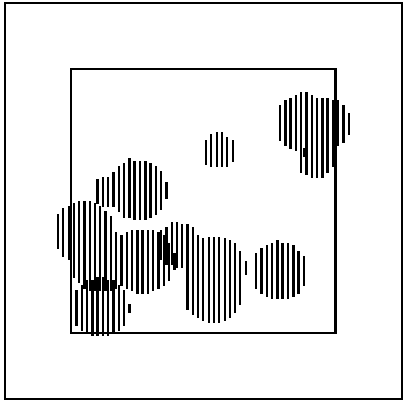
The cross section of the system at $t=1.5$ for $\eta_{tot} = 50$.

\end{center}

}

\pagebreak

%

{\it
\begin{center}
Fig. 6

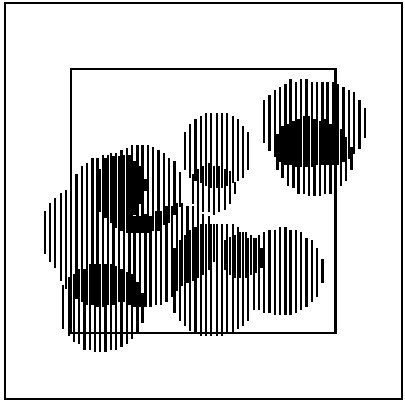
The cross section of the system at $t=3$ for $\eta_{tot} = 50$.

\end{center}

}

\pagebreak

%

{\it
\begin{center}
Fig. 7

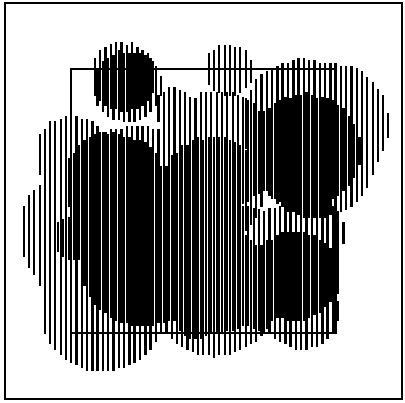
The cross section of the system at $t=6$ for $\eta_{tot} = 50$.

\end{center}

}

\pagebreak

%

{\it
\begin{center}
Fig. 8

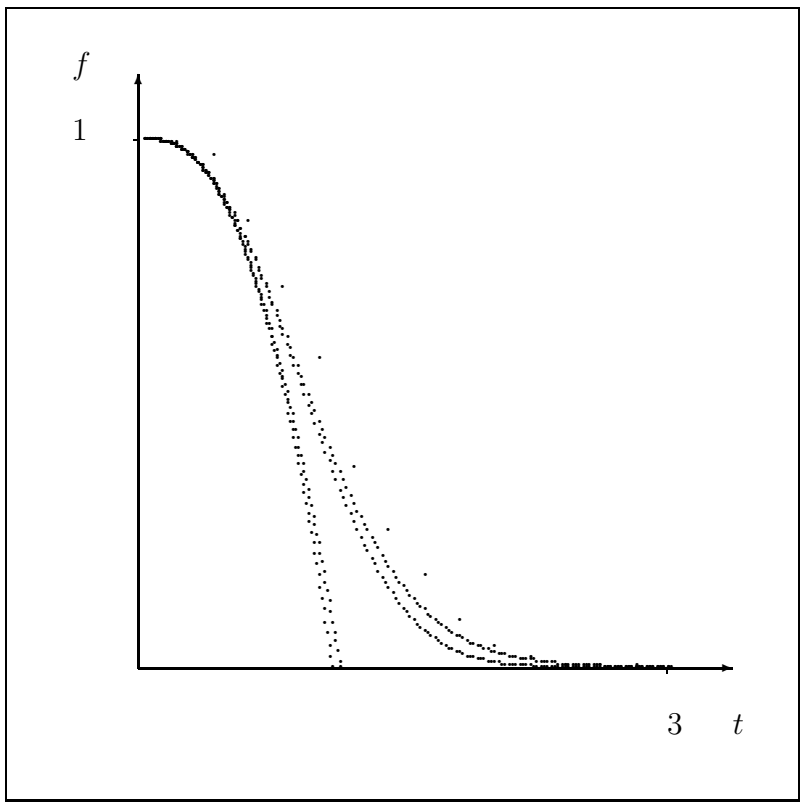
Comparison of different models in the quasihomogeneous situation.

\end{center}

}

\pagebreak

%

{\it
\begin{center}
Fig. 9

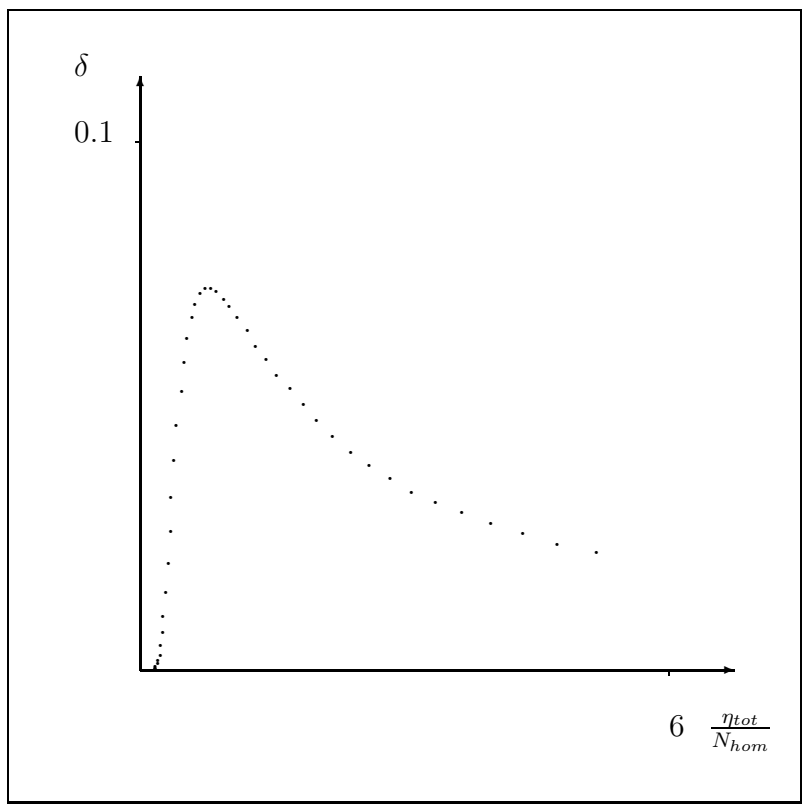
Relative error of the quasihomogeneous approach in the first model.

\end{center}

}

\pagebreak

%

{\it
\begin{center}
Fig. 10

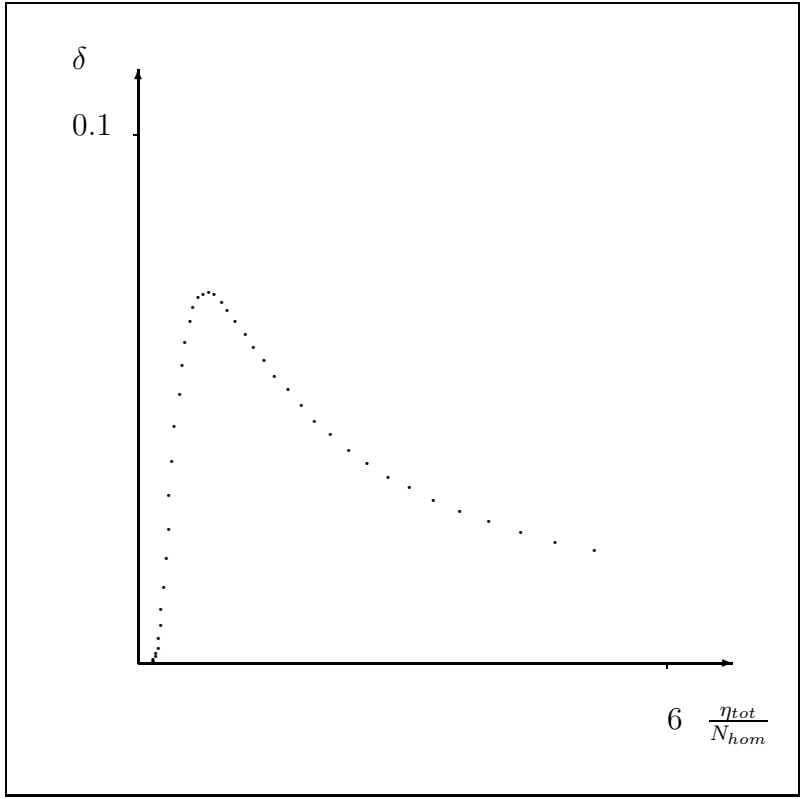
Relative error of the quasihomogeneous approach in the second model.

\end{center}

}

\pagebreak

%

{\it
\begin{center}
Fig. 11

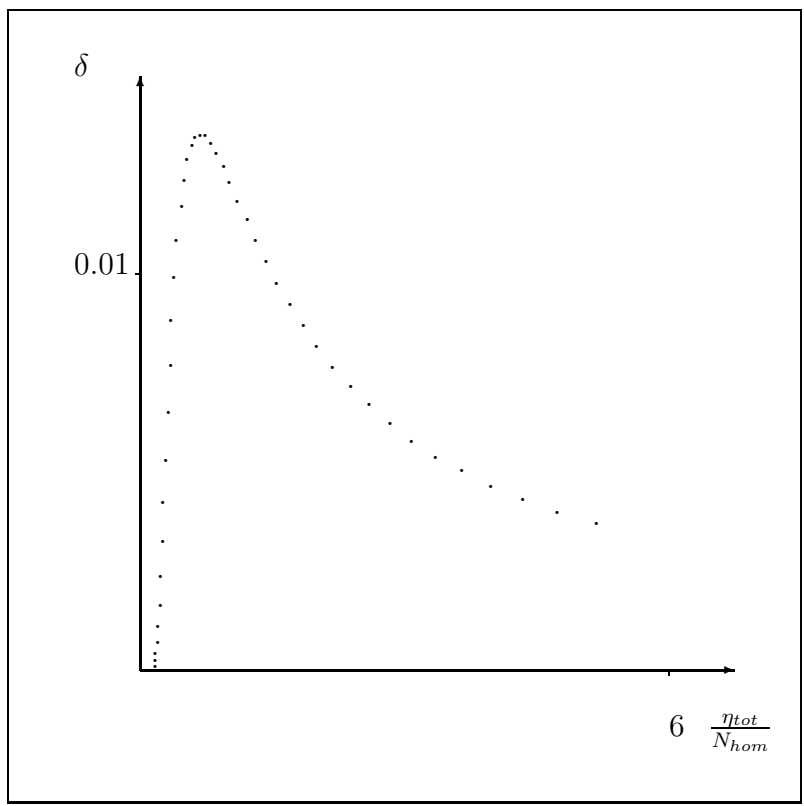
Relative error of the quasihomogeneous approach in the third model.

\end{center}

}

\pagebreak

%

{\it
\begin{center}
Fig. 12

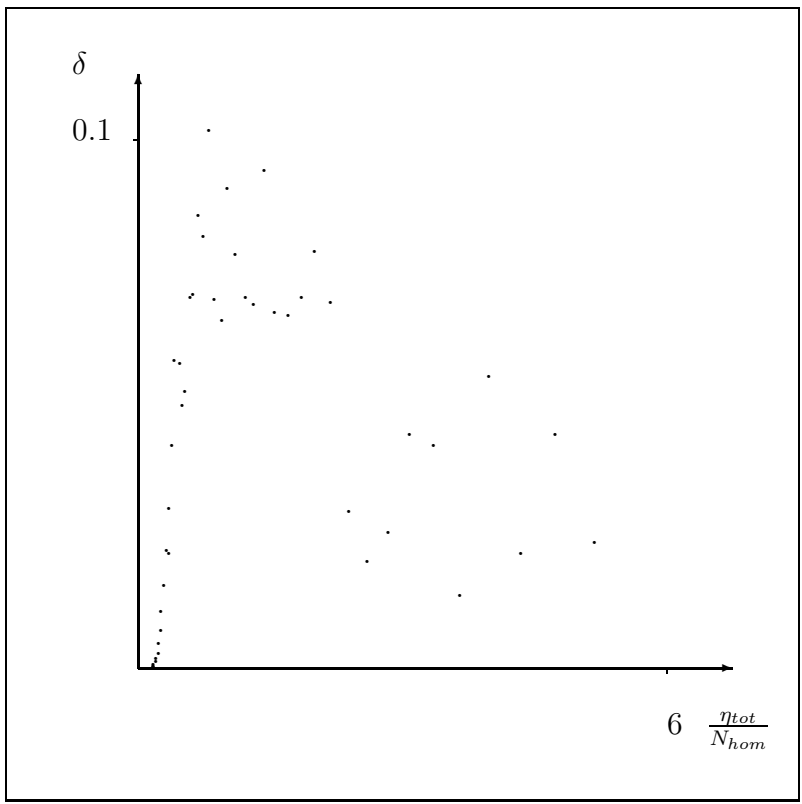
Relative error of the quasihomogeneous approach in the universal simulation.

\end{center}

}

\end{document}